# YBCO single grains seeded by 45° – 45° bridge-seeds of different lengths

Y-H Shi, J H Durrell, A R Dennis and D A Cardwell

Bulk Superconductivity Group, Engineering Department, University of Cambridge, CB2 1PZ, UK

E-mail: ys206@cam.ac.uk

**Abstract**. Single grain, (RE)BCO bulk superconductors in large or complicated geometries are required for a variety of potential applications, such as motors and generators and magnetic shielding devices. As a result, top, multi-seeded, melt growth (TMSMG) has been investigated over the past two years in an attempt to enlarge the size of (RE)BCO single grains specifically for such applications. Of these multi-seeding techniques, so-called bridge seeding provides the best alignment of two seeds in a single grain growth process. Here we report, for the first time, the successful growth of YBCO using a special, 45° – 45°, arrangement of bridge-seeds. The superconducting properties, including trapped field, of the multi-seeded YBCO grains have been measured for different bridge lengths of the 45° – 45° bridge-seeds. The boundaries at the impinging growth front and the growth features of the top, multi-seeded surface and cross-section of the multi-seeded, samples have been analysed using optical microscopy. The results suggest that an impurity-free boundary between the two seeds of each leg of the bridge-seed can form when 45° – 45° bridge-seeds are used to enlarge the size of YBCO grains.

## 1. Introduction

Single grain, (RE)BCO high temperature superconductors (HTS) have potential for a variety of engineering applications due to their ability to trap magnetic fields much greater than those that can be achieved with iron-based permanent magnets. The world record trapped field for a stack of two YBCO single grains, each of diameter 25 mm, is 17 T at 29 K [1]. A practical limitation to the manufacture of these samples, however, is their relatively low growth rate, with a single grain of diameter 30 mm taking around one week to grow. As a result, reducing the processing time of single grain bulk HTS is fundamental to developing a practical, industrial growth process. Multi-seeding is one technique that can be used potentially to enlarge the size of (RE)BCO single grains and reduce processing time. In addition, the multi-seeding process provides an opportunity to study the grain boundaries in the bulk material, which are physically different to those that form in thin films and tapes due primarily to the presence of RE-211 particles in bulk material fabricated by top seeded melt growth (TSMG).

Multi-seeding using two or more separate seeds has been used previously to enlarge the size of (RE)BCO single grains [2-9]. In most of these studies, however, poorly connected grain boundaries have characteristically formed between the two seeds, characterized by a build-up of impurity phases [2-5]. Trapped field, which is considered generally to be the most important superconducting property for practical applications, was observed to be low in the small number of cases where it was measured in these samples [6], compared to the trapped field generated typically by a single grain grown from a single-seed. We have developed a novel, bridge-seeding technique to improve significantly the alignment between seeds in the top multi-seeded melt growth (TMSMG) process [10][11], which minimises any misinterpretation of the properties of the multi-seeded sample that may be associated with seed misalignment, rather than with grain growth. We reported recently that the effect of bridge length of 0° – 0° orientated bridge-seeds on trapped field and $J_c$ and concluded that such an alignment of seeds is effective for the enlargement of single (RE)BCO grains for 0° – 0° bridge-seeds of length up to 8 mm [12]. However, grain boundaries containing impurities (such as solidified liquid phases) still exist in these samples at the interface of the grains grown from the two legs of the bridge-seeds. In this paper a special 45° – 45° arrangement of bridge-seeds is used to multi-seed successfully YBCO single grains. The effects of length of the 45° – 45° bridge-seeds on the trapped field are discussed and the growth related features, grain microstructures and grain boundaries generated by the multi-seeding process are analysed.

## 2. Experimental

### 2.1. Fabrication of the 45° – 45° bridge-seeds

Initially, SmBCO single grains (as shown schematically in figure 1(a)) were fabricated by TSMG using generic seeds [13, 14] from a mixed precursor powder of composition 75wt% Sm123 + 25 wt% Sm211 +0.1wt% Pt. The as-grown single grains were cut into slices of length 4 mm, 7 mm and 10 mm along a facet line (indicated by the dashed red lines in figure 1(a)) on the top surface of the grain (i.e. corresponding to the (011) direction in the unoxygenated, tetragonal lattice). These slices were ground into bridge shaped seeds, as shown in figure 1(b). These so-called 45° – 45°, bridge-seeds, labeled according to their relative orientation in the parent grain, were used successfully to multi-seed the bulk YBCO samples due primarily to the controlled alignment and relative orientation of the two legs of the seed.

### 2.2. Melt-processing single grains using 45-45 bridges in different length

Y-123 (99.9%), Y-211 (99.9%) and $CeO_2$ (99.9%) commercial powders were mixed in a composition ratio of 70wt% Y-123 + 30wt% Y-211 + 1wt% $CeO_2$ using an electric pestle and mortar. The mixed powders were pressed uniaxially into pellets of diameter 32 mm in diameter and height 15 mm. These pellets were then pressed in a cold isostatic press under a pressure of 180 MPa. The 45° – 45° bridge-seeds were placed on the top surface of the pressed pellets, which, in turn, were placed in a large chamber furnace and melt-processed using TMSMG technique. Bridge-seeds of lengths 0 mm (i.e. a single seed), 4 mm, 7 mm and 10 mm were used to melt process the samples, respectively. The heating profile used to fabricate the grains involved increasing the temperature initially at relatively high rate of 150 °C/hour to 1045 °C, holding for 1 hour, cooling quickly to 1005 °C and cooling slowly at 0.3°C/hour to 972°C. Finally, the samples were furnace-cooled to room temperature. Two batches, each containing 4 samples grown from bridge-seeds with separation distances of 0, 4, 7 and 10 mm, of diameter 25 mm and height 12 mm were grown in this research, as shown in figure 2. Finally, the as-grown samples were oxygenated in flowing oxygen at 420 °C for one week.

## 2.3 Trapped field measurements and microstructure analysis

The top surfaces of the 8 samples were polished flat for trapped field measurement. Each sample was field cooled to 77 K using liquid nitrogen in 1.5 T applied perpendicular to its top surface. The applied field was then removed and the trapped field on the top surface of each sample measured using a rotating array of 20 Hall probes. The distance between the sample surface and the Hall probes was estimated to be 0.7 mm. Each sample was then top-sliced by 1.5 ± 0.3 mm using a diamond edge cutting disc, polished, and the trapped field re-measured effectively at a different sample depth. Each sample in a second set (5–8, one grown with each seed length) was cut parallel to its thickness to expose a cross section for microstructural analysis.

## 3. Results and discussion

### 3.1. Trapped field at the top surface of the as-grown samples

Figures 3(a) and (b) show photographs of the polished top surfaces and trapped field profiles of samples 1–4. The colour contour scale for the trapped field is the same for each sample in figure 3(b). It can be seen from this figure that the 45° – 45° bridge-seeds multi-seed YBCO single grains very effectively, and that bridge-seeds of length up to 7 mm can be used to grow a single grain with a single peak in the trapped field profile. A 10 mm 45-45 bridge seed in this research still can grow a YBCO grain with a trapped field peak value 0.75 T, which is relatively high for samples of diameter 25 mm, and is comparable to the trapped field generated typically by a single-seeded YBCO sample of the same size. It can be seen that the trapped field profile of sample 4 exhibits two peaks. These peaks, however, are relatively indistinct, and are not separated by a deep minimum. As a result, the trapped field at the centre of the sample remains relatively high at around 0.45 T, unlike YBCO fabricated using a 0° – 0° bridge-seed of length 10 mm, which produces two separate and very distinct peaks in the trapped field, as reported previously [12]. On closer inspection, a thin crack between the two legs of the bridge-seed is apparent on the top surface of sample 4, which explains why the trapped field profile exhibits a double peak structure. This, in turn, suggests that the crack on the top surface of sample 4 does not extend throughout the entire thickness of the sample (i.e. it is a relatively "shallow" crack). This crack appears to be caused by misalignment between the grains at the point of impingement, which is apparent from the mismatch of the two facet lines (indicated by yellow arrows) in the polished top surface of sample 4 (figure 3(a)). Significantly, no second phase (such as solidified liquid phase) was observed by high-resolution optical microscopy to accumulate at the crack, such as that present typically in 0° – 0° bridge-seeded samples [12]. This suggests that the grain boundary formed using 45° – 45° bridge-seeds of length up to 10 mm is still relatively clean and strongly connected.

Figure 4 shows the distribution of peak values of trapped field as a function of 45° – 45° bridge-seed length. It can be seen that no particular trend in these data is apparent, and that the grain boundaries formed in melt processed YBCO samples are less sensitive to bridge-seed length than samples grown using 0° – 0° bridge-seeds. This suggests that the field trapping capability of single grains grown using bridge-seeds is determined by both the length of the bridge and by the orientation of the seed legs. In other words, it is unlikely that the relative orientation of the 45° – 45° bridge-seeds is exactly 45°, and that a misalignment of up to 2° is likely given that the seed was cut from the parent sample by eye using the facet line as a guide. As a result, the bridge-seeds can be classified as 45°±θ – 45°±θ, where $0 > \theta > 2°$.

Figure 3(a) shows the features of the polished top surfaces of samples 1–4. The boundaries of the *ab* and *c* growth sectors are defined by the square structures at the position of the seeds for each sample, with the *c* growth sector located within the square. The appearance of sample 2 is similar to that of sample 1, which suggests that short, 45° – 45° bridge-seeds behave effectively as a larger, single seed, but with a wider *c* growth sector than that observed for a single-seeded sample. Figure 5 shows photographs of the cross sections of samples 6, 7 and 8 seeded using 45° – 45° bridge-seeds of length 5, 7 and 10 mm. It can be seen that the boundaries of *ab* and *c* growth sectors of the sample seeded by the bridge-seed of length 5 mm are similar to those of the sample seeded by a single seed (sample 1) and that only the apparent seeding area is larger. This figure provides further evidence that 45° – 45° bridge-seeds of length between 4 and 5 mm behave effectively as a large, single seed.

Two *c* growth sectors are evident in the top surface of the sample when bridge-seeds of length 7 mm and 10 mm are used (figure 3(a)). These sectors join together at a later stage in the growth process to form one *c* growth sector at the lower part of each sample, as shown in figure 5. It can be predicted from the growth sector features evident from the sample cross-sections that the lower part of the grains seeded by longer, 45° – 45° bridge-seeds (7 and 10 mm in this case) are similar to those observed in single grains melt processed using a single-seed. In addition, the length of the 45° – 45° bridge-seed correlates directly with the width of the complete *c* growth sector. Although a similar correlation is observed for samples fabricated using 0° – 0° bridge-seeds, no vertical grain boundaries are present in the cross-sections of samples fabricated using the 45° – 45° bridge-seeds. The cracks labeled A, B and C in figure 5 extend either diagonally through the sample cross section (cracks A and C) or parallel to the *ab* plane (crack B). In addition, there is no second phase, such as a solidified liquid phase, between the two seeds of the bridge in the samples seeded using longer bridge-seeds. This must be because the growth fronts meet perpendicularly for the 45° – 45° bridge-seed, in a similar way to which the facet lines form on the top surface of a single-seeded sample to form a strongly connected region. The 45° angle of grain impingement results in the pushing of impurity phases towards the edge of the sample, along the boundary. This differs to the case of multi-seeding using 0° – 0° bridge-seeds, which results in the trapping of impurities at the mid-point of the impinging grains (i.e. there is no pushing force on the impurity phase along the boundary).

It is clear that the 45° – 45° bridge-seeds can be used to grow grains with higher trapped fields than 0° – 0° bridge-seeds in the previous study [12]. It should be noted that the Y-123 precursor powder used to form the precursor pellets in the 0° – 0° bridge-seed study was obtained from a different manufacturer than that used in the current work, although this can not explain the considerable improvement in field trapping properties and microstructural features of the samples seeded with the 45° – 45° bridge-seeds. The potential reduction in fabrication time by the use of the multi-seeding technique was not evaluated in this study, since the samples were batched processed to save time.

### 3.2. Trapped field at the top surface of the top-sliced samples

Figures 6(a) and (b) show the polished top surfaces and trapped fields of samples (1–4) after the removal of a slice of thickness 1.5 mm from the top surface of the as-grown samples. It can be seen that the trapped field profiles become smoother compared to those in figure 3(b). This supports the observation that the microstructural features of the samples grown using the 45° – 45° bridge-seeds evolve gradually towards those of samples grown using single seeds through the sample thickness (i.e. as the mismatch caused by the seeding process decreases).

Finally, it is apparent that the trapped field varies from sample to sample, even for the same bridge-seed length. Evidence has been presented that seed alignment is critical to the formation of a strongly connected grain boundary. Although the length of the bridge-seed can be varied accurately, it is likely that the orientation of each bridge will be exactly 45° – 45°, which may vary randomly over a small range, and reasonably by ±2°. The highest trapped fields were observed in the present study in samples fabricated using the best-aligned bridge-seeds.

## 4. Conclusions

45° – 45° SmBCO bridge-seeds have been used to grow YBCO single grains successfully. A bridge-seed of length up to 7 mm can grow YBCO single grains that have microstructural features that are similar to grains grown with a single-seed. There is no sharp decrease in trapped field at the centre of YBCO grains fabricated using a bridge-seed of length up to at least 10 mm, as is observed for samples fabricated using 0° – 0° bride-seeds. A peak trapped field value of 0.75 T has been achieved in this study for YBCO seeded by a 45° – 45° bridge-seed of length 10 mm. The alignment of seeds in the multi-seeding process is critical to the formation of a strongly connected grain boundary. Bridge-seeds can provide the best alignment of two seeds reported to date in any multi-seeding process. The trapped fields produced by samples fabricated from 45° – 45° bridge-seeds are more tolerant to the length of the bridges compared to those fabricated using 0° – 0° bridge-seeds. An impurity-free boundary between the two seeds of the bridge can form when 45° – 45° bridge-seeds are used to enlarge YBCO single grains. Short, 45° – 45° bridge-seeds behave as one large single seed, whereas longer bridge-seeds may result in the formation of grain boundaries that are relatively shallow and clean, unlike the case of boundaries formed using 0° – 0° bridge-seeds.

**Figure captions**

| | |
|---|---|
| Figure 1 | Schematic diagrams showing how the 45° – 45° bridge-seeds were fabricated. (a) Top view of a single grain and (b) the 45° – 45° bridge-shaped seed. |
| Figure 2 | Melt-processed YBCO grains (1-4) fabricated using 45° – 45° bridge-seeds of length 0 mm, 4 mm, 7mm and 10 mm, respectively |
| Figure 3(a) | Photographs of the polished top surfaces of one set of YBCO grains (1-4) seeded by 45° – 45° bridge-seeds of length 0, 4, 7 and 10 mm, respectively |
| Figure 3(b) | Trapped field at the top surfaces of the samples in figure 3(a). |
| Figure 4 | Distribution of the peaks trapped fields for samples 1-8. |
| Figure 5 | Photographs of the cross-sections through the thickness of samples 5, 6, 7 and 8 |
| Figure 6(a) | Photographs of the polished top surfaces after top-slicing 1.5 mm from the as-grown YBCO samples (1-4) grown from 45° – 45° bridge-seeds of length 0, 4, 7 and 10 mm. |
| Figure 6(b) | Trapped fields of the samples in figure 6(a). |

Figure 1

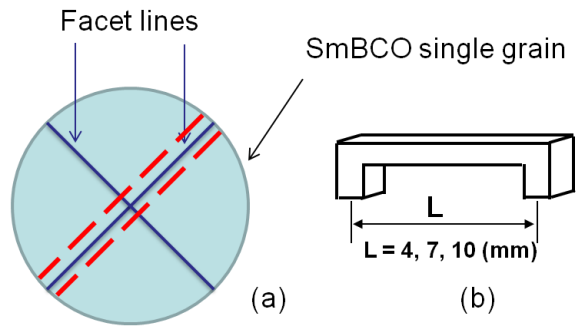

Figure 1 Schematic diagrams showing how the 45° – 45°, bridge-seeds were fabricated. (a) Top view of the SmBCO single grain and (b) A 45° – 45° bridge shaped seed.

Figure 2

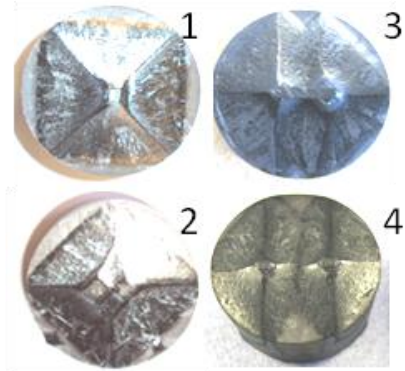

Figure 2 Melt-processed YBCO grains (1–4) grown by TMSMG using 45° – 45° bridge-seeds of leg separation length 0 mm, 4 mm, 7mm and 10 mm, respectively.

Figure 3

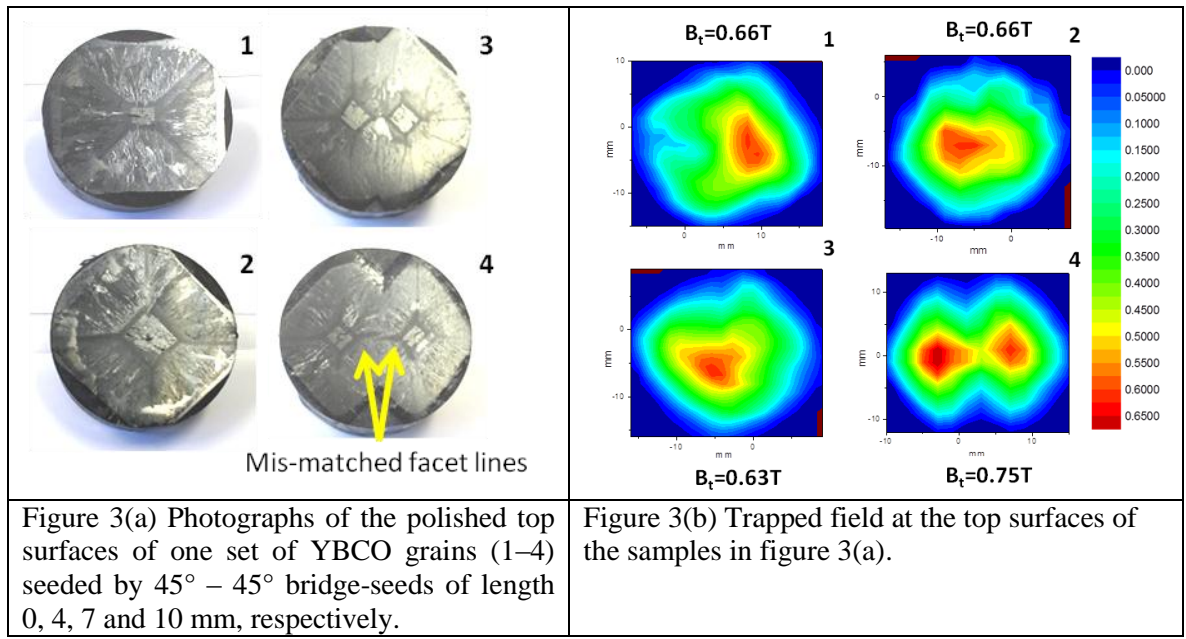

| Figure 3(a) Photographs of the polished top surfaces of one set of YBCO grains (1–4) seeded by 45° – 45° bridge-seeds of length 0, 4, 7 and 10 mm, respectively. | Figure 3(b) Trapped field at the top surfaces of the samples in figure 3(a). |

Figure 4

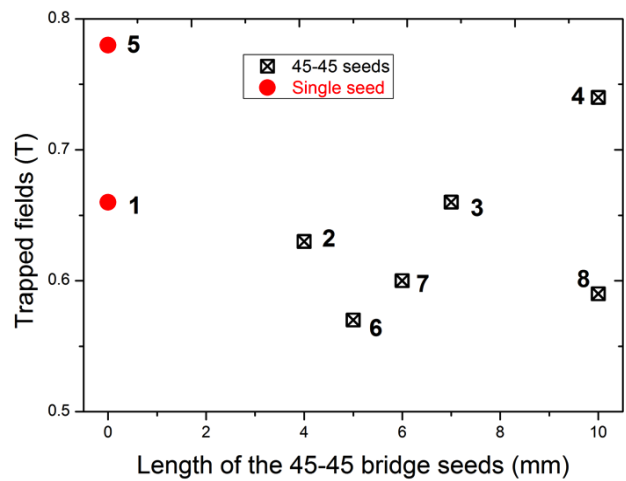

Figure 4 Distribution of the peak trapped field for samples 1-8.

Figure 5

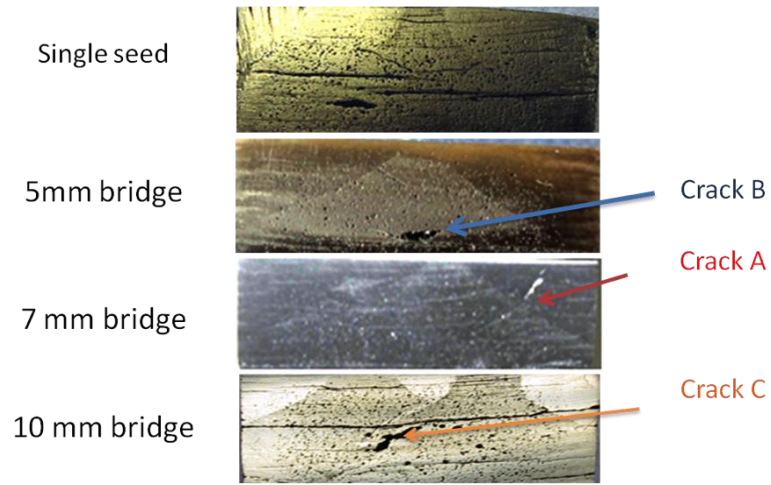

Figure 5 Photographs of the cross-sections through the thickness of samples 5, 6, 7 and 8.

Figure 6

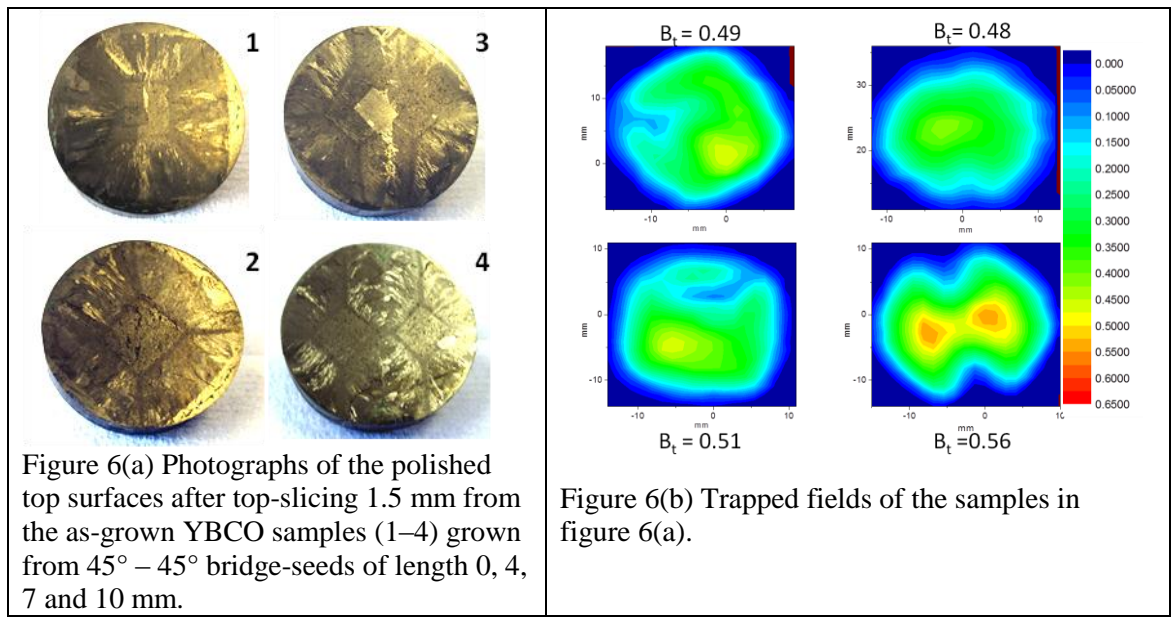

Figure 6(a) Photographs of the polished top surfaces after top-slicing 1.5 mm from the as-grown YBCO samples (1–4) grown from 45° – 45° bridge-seeds of length 0, 4, 7 and 10 mm.

Figure 6(b) Trapped fields of the samples in figure 6(a).